\documentclass[a4paper,10pt,twoside]{cpc-hepnp}

\usepackage{multicol}
\usepackage{graphicx}
\usepackage{booktabs}
\usepackage{amssymb,bm,mathrsfs,bbm,amscd}
\usepackage[tbtags]{amsmath}
\usepackage{lastpage}
\usepackage{xcolor}

\newcommand{\beqa}{\begin{eqnarray}}
\newcommand{\eeqa}{\end{eqnarray}}
\newcommand{\beq}{\begin{equation}}
\newcommand{\eeq}{\end{equation}}

\begin{document}

\fancyhead[co]{\footnotesize R. Williams \emph{et al}:Bethe-Salpeter equations: mesons beyond the rainbow-ladder truncation}

\footnotetext[0]{Received 15 December 2009}

\title{Bethe-Salpeter equations: mesons beyond the rainbow-ladder truncation}

\author{%
      Richard Williams$^{1;1)}$\email{richard.williams@physik.tu-darmstadt.de}%
\quad Christian S. Fischer$^{1,2}$
}
\maketitle

\address{%
1~(Institute for Nuclear Physics, 
 Darmstadt University of Technology, 
 Schlossgartenstra{\ss}e 9, 64289 Darmstadt, Germany)\\
2~(GSI Helmholtzzentrum f\"ur Schwerionenforschung GmbH, 
  Planckstr. 1  D-64291 Darmstadt, Germany.)\\
}

\begin{abstract}
We investigate masses of light mesons from
a coupled system of Dyson--Schwinger (DSE) and Bethe--Salpeter equations
(BSE), taking into account dominant non-Abelian, sub-leading Abelian, and
dominant pion cloud contributions to the dressed quark-gluon vertex. 
The  axial-vector Ward-Takahashi identity preserving Bethe-Salpeter 
kernel is constructed and the spectrum of light mesons calculated. Our model goes
significantly beyond the rainbow-ladder.  We find
that sub-leading Abelian corrections are further dynamically suppressed,
and that our results supersede early qualitative predictions from
simple truncation schemes. 
\end{abstract}



\begin{multicols}{2}

\section{Introduction}\label{intro}
 Using the Dyson-Schwinger and Bethe-Salpeter equations, there
have been successful descriptions of pseudoscalar and
vector
mesons~\cite{Maris:1997hd,Maris:1997tm,Maris:1999nt,Maris:2000sk,Bhagwat:2006pu}
using the simple rainbow-ladder (RL) truncation. This is not unexpected since
for the spectrum of light pseudoscalar mesons, it is the axial-vector
Ward-Takahashi identity that governs the pattern of dynamical symmetry
breaking.  These RL investigations have also been recently 
extended to describe baryons~\cite{Nicmorus:2008vb,Eichmann:2009qa}; a
status report is given in these proceedings~\cite{Alkofer:2009jk}.

Considerable effort to go beyond RL, using the prescription
of~\cite{Munczek:1994zz,Bender:1996bb}, has been made in
a number of works~\cite{Bender:1996bb,Bender:2002as,Bhagwat:2004hn,Watson:2004kd,Matevosyan:2006bk}.
These however considered sub-leading Abelian corrections rather than the
dominant non-Abelian parts of the quark-gluon vertex~\cite{Alkofer:2008tt}. 
In Ref.~\cite{Fischer:2009jm} we performed a first calculation of the
leading non-Abelian diagrams in a coupled DSE/BSE approach. Here we include 
the sub-leading Abelian corrections together with
the pion back-reaction~\cite{Fischer:2008wy,Williams:2009wx}.

\section{Dyson-Schwinger Equations}\label{sec:dse}

To solve the DSEs for the quark propagator and quark-gluon vertex, shown
in Fig.~\ref{fig:qgdse}(a) and Fig.~\ref{fig:qgdse}(b) we must introduce 
a truncation scheme. The BSE
must be consistently truncated, being careful to preserve chiral symmetry.
This is expressed via the axial-vector
Ward-Takahashi identity (axWTI) and ensures that pions are 
the pseudo-Goldstone bosons.

The
simplest truncation that satisfies this criterion is that of
RL whereby the full quark-gluon vertex is replaced by a
bare vertex~\cite{Maris:1997tm}. The axWTI preserving 
kernel in the BSE then corresponds to a single gluon exchange,
re-summed to all orders thus providing the `ladder'. These 
RL models effectively subsume additional vertex corrections from the 
Yang-Mills sector and unquenching via an effective coupling. 

To make an extension to the rainbow-ladder truncation, we investigate
the quark-gluon vertex and determine the impact of corrections beyond
tree-level to our quarks and mesons~\cite{Bender:1996bb,Bender:2002as,Bhagwat:2004hn,Watson:2004kd,Bhagwat:2004kj,Matevosyan:2006bk,Chang:2009zb}.
Following the analysis of~\cite{Alkofer:2008tt,Fischer:2007ze} we approximate the full
DSE Fig.~\ref{fig:qgdse}(a) with the (nonperturbative) one-loop structure of
Fig.~\ref{fig:qgdse}(b). Here the first
`non-Abelian' loop-diagram in Fig.~\ref{fig:qgdse}(b) subsumes the first two diagrams in the full
DSE to first order in a skeleton expansion of the four-point functions.
We neglect the two-loop diagram in the full DSE Fig.~\ref{fig:qgdse}(a), which is
justified for small and large momenta~\cite{Alkofer:2008tt,Bhagwat:2004kj}. 
The remaining `Abelian' contributions are split
into the non-resonant second loop-diagram in Fig.~\ref{fig:qgdse}(b) and a third
diagram containing effects due to hadron back-reactions. The details of
including the pion-backreaction are contained within
Refs.~\cite{Fischer:2007ze,Fischer:2008sp,Fischer:2008wy}, see therein
for details. This provides an extension of our previous truncation
involving only the non-Abelian diagram~\cite{Fischer:2009jm}. 

\end{multicols}

\begin{center}
\begin{eqnarray}
\mathrm{(a)}\begin{array}{c}
  \includegraphics[width=0.66\columnwidth]{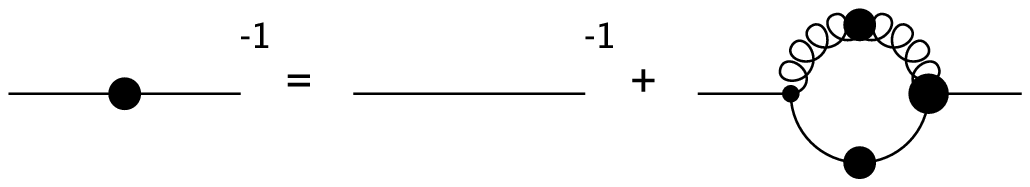}
\end{array}\label{fig:quarkdse}\nonumber\\
\mathrm{(b)}\begin{array}{c}
  \includegraphics[width=0.71\columnwidth]{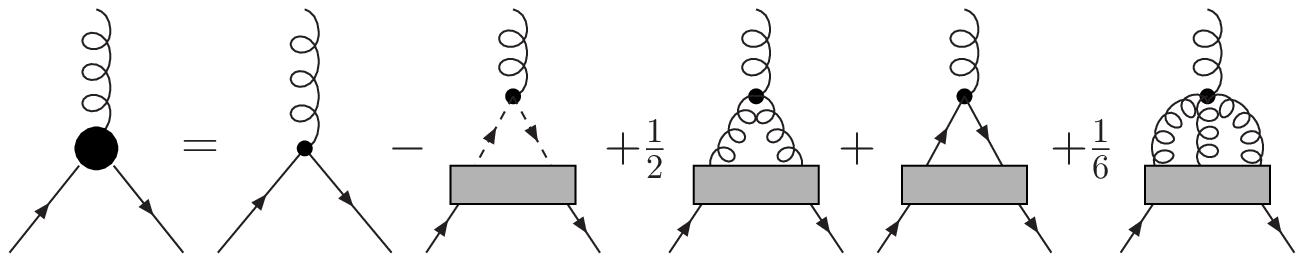}
\end{array}\label{fig:dse1}\nonumber\\
\mathrm{(c)}\begin{array}{c}
  \includegraphics[width=0.71\columnwidth]{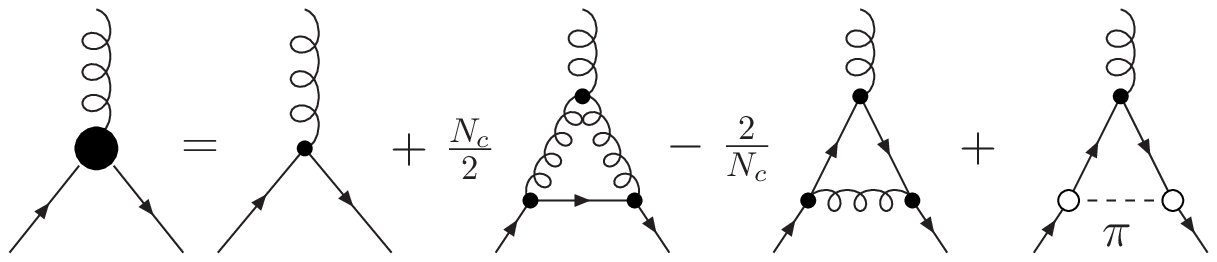}
\end{array}\label{fig:dse2}\nonumber
\nonumber
\end{eqnarray}
\figcaption{
DSEs for: (a) fully dressed quark propagator;
(b) full quark-gluon vertex;
(c) truncated quark-gluon vertex. Internal
propagators are dressed, with gluons shown by wiggly lines, quarks by
straight lines and dashed lines mesons. White-filled circles show
meson amplitudes whilst black-filled represent vertex dressings.\label{fig:qgdse}}
\end{center}

\begin{multicols}{2}

\section{Bethe-Salpeter equation}\label{sec:bse}
The Bethe-Salpeter equation describing a relativistic bound-state of mass $M$ is
calculated through
\begin{equation}
  \left[\Gamma(p;P)\right]_{tu} = \lambda\! \int_k
K_{tu}^{rs}(p,k;P)\left[S(k_+)\Gamma(k;P)S(k_-)\right]_{sr}\,.
\label{eqn:bse2}
\end{equation}
Here $\int_k=\int \frac{d^4 k}{\left( 2\pi \right)^4}$, $\Gamma(p;P) \equiv \Gamma^{(\mu)}(p;P)$ is the Bethe-Salpeter
vertex function of a quark-antiquark bound state. It is
a homogeneous equation, with a discrete spectrum of solutions at momenta
$P^2=-M_i^2$ corresponding to $\lambda\left(P^2\right)=1$. The lightest
of these $M_i$ pertains to the ground state solution. The momenta $k_+ =
k +\eta P$ and $k_- = k-(1-\eta)P$ are such that the total momentum $P$
of the meson is given by $P=k_+-k_-$ and the relative momentum $k=\left(
k_+ +k_- \right)/2$. The momentum partitioning parameter is $\eta$,
of which physical observables are independent. For convenience we choose
it to be $1/2$ without loss of generalisation. The object $K_{tu}^{rs}(p,k;P)$ is the Bethe-Salpeter
kernel, whose Latin indices refer to colour, flavour and Dirac
structure. 

The Bethe-Salpeter vertex function $\Gamma^{(\mu)}(p;P)$ can be 
decomposed into eight Lorentz and Dirac structures.  The structure is 
constrained by the transformation properties under CPT of the meson~\cite{LlewellynSmith:1969az}. 

The consequences of chiral symmetry can be expressed through the
axial-vector Ward-Takahashi identity for flavour non-singlet mesons
\begin{equation}
  -i P_\mu \Gamma^5_\mu = S_F^{-1}(p_+)\gamma_5 + \gamma_5
  S_F^{-1}(p_-)-2m_R \Gamma^5(p;P)\;,
  \label{eqn:avwti}
\end{equation}
with $p$ and $P$ the relative and total momentum of the meson
respectively, $p_\pm$ is combination $p\pm P/2$, $\Gamma^5_\mu$ the axial-vector and $\Gamma^5$ the
pseudoscalar vertex.

\end{multicols}
\ruleup
\begin{center}
	\begin{eqnarray}
	  \begin{array}{c}
		\includegraphics[scale = 0.50]{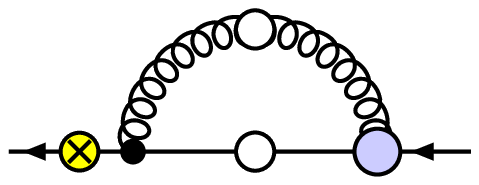}
	  \end{array}
	  \!+\!
	  \begin{array}{c}
		\includegraphics[scale = 0.50]{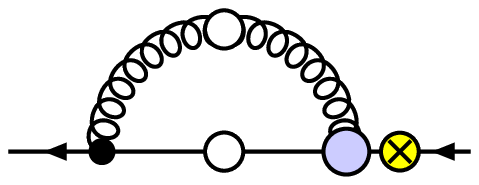}
	  \end{array}
	  =
	  -\!\!\!
	  \begin{array}{c}
		\includegraphics[scale = 0.50]{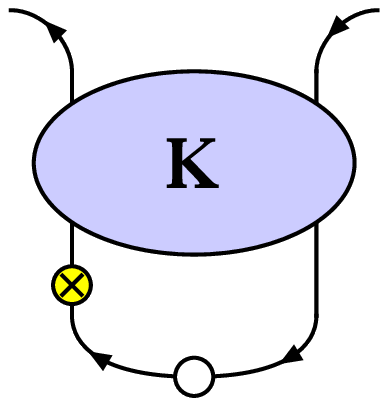}
	  \end{array}
	  \!-\!\!
	  \begin{array}{c}
		\includegraphics[scale = 0.50]{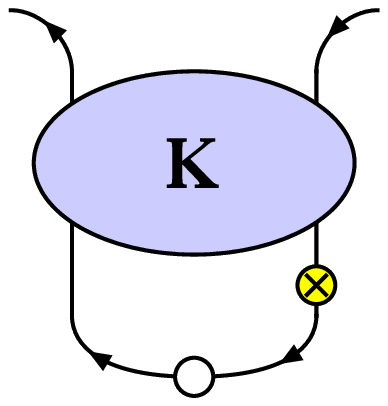}
	  \end{array}\nonumber
	\end{eqnarray}
\figcaption{The flavour non-singlet axWTI written showing the connection
between the quark self-energy and Bethe-Salpeter kernel. All
propagators are dressed, with wiggly and straight lines for gluons
and quarks respectively. The crossed circle indicates a $\gamma_5$
insertion.\label{fig:axwti}}
\end{center}
\begin{multicols}{2}

It is well-known that one may construct a Bethe-Salpeter kernel $K_{tu}^{rs}$
satisfying the axWTI by means of a functional derivative of the quark 
self-energy. Applying this cutting procedure to the quark DSE of
Fig.~\ref{fig:qgdse}(a) with the quark-gluon vertex of Fig.~\ref{fig:qgdse}(c)
yields the Bethe-Salpeter equation portrayed in Fig.~\ref{fig:ourbse}~\cite{Maris:2005tt}.
Because we do not trivialize the momentum
dependence of our propagators, the calculation is genuinely two-loop. Such a truncation allows us to consider the dominant non-Abelian
corrections~\cite{Fischer:2009jm}, together with the sub-leading Abelian 
corrections~\cite{Watson:2004kd}. 
We are now in a position to perform a systematic comparison of the
impact of each vertex correction on the spectrum of light mesons. 
\end{multicols}
\ruleup

\begin{center}
\begin{eqnarray*}
	\begin{array}{c}
	\includegraphics[scale=0.7]{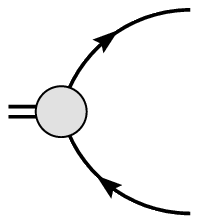}\\
	\end{array}
	&=& 
	\begin{array}{c}
	\includegraphics[scale=0.7]{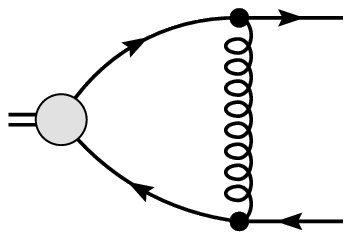}\\
	\end{array}
	+
	\begin{array}{c}
	\includegraphics[scale=0.7]{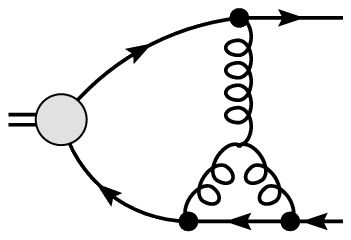}\\
	\end{array}
	+
	\begin{array}{c}
	\includegraphics[scale=0.7]{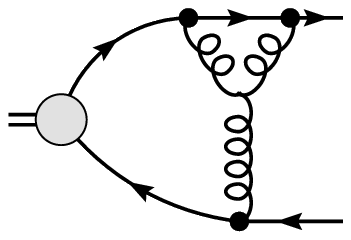}\\
	\end{array}
	+
	\begin{array}{c}
	\includegraphics[scale=0.7]{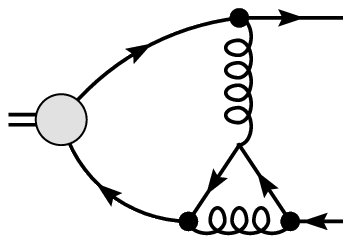}\\
	\end{array}\nonumber\\
	&+&
	\begin{array}{c}
	\includegraphics[scale=0.7]{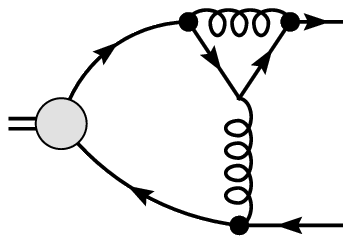}\\
	\end{array}
	+
	\begin{array}{c}
	\includegraphics[scale=0.7]{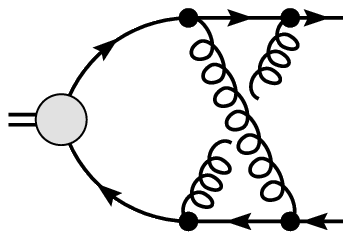}\\
	\end{array}
	+
	\begin{array}{c}
	\includegraphics[scale=0.7]{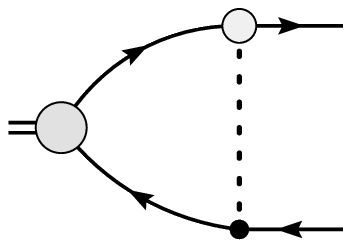}\\
	\end{array}
	+
	\begin{array}{c}
	\includegraphics[scale=0.7]{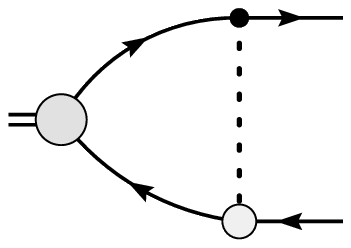}\\
	\end{array}
	\nonumber
	\end{eqnarray*}
\figcaption{The axWTI preserving BSE corresponding to our vertex truncation. All 
propagators are dressed, with wiggly and straight lines showing gluons
and quarks respectively.\label{fig:ourbse}}
\end{center}
\ruledown
\begin{multicols}{2}

\section{Results}\label{sec:results}

\begin{center}
\renewcommand{\arraystretch}{1.1} 
\begin{tabular}{@{}c||ccccc}
  \toprule Model & $m_\pi$ &  $m_\sigma$ & $m_\rho$  & $m_{a_1}$ & $m_{b_1}$\\
  \hline
  RL      & $138$  &  $645$  & $758$ & $926$ & $912$\\
  NA      & $142$  &  $884$  & $881$ & $1056$ & $973$\\
  AB      & $137$  &  $602$  & $734$ & $889$ & $915$\\
  AB+NA   & $142$  &  $883$  & $878$ & $1052$ & $972$\\
  NA+PI   & $138$  &  $820$  & $805$ & $1040$ & $941$\\
  \hline
PDG~\cite{Amsler:2008zzb}   & $138$  & $400$--$1200$ & $776$ & $1230$ & $1230$ \\
\bottomrule
\end{tabular}
\tabcaption{Masses for a variety of mesons calculated
using rainbow-ladder (RL), additional
corrections from the Abelian (AB) and non-Abelian (NA) diagrams, and
with the pion backreaction (PI). Masses are given in MeV. \label{tab:results}}
\end{center}

We present results in Table~\ref{tab:results}. The benchmark for
comparison,
for which the parameters of the interaction are fixed to meson
observables, are those of the rainbow-ladder (RL) approximation. By
including the dominant non-Abelian vertex correction (NA) we find only a
small change in the pion mass, as expected since it is a
pseudo-Goldstone boson. For the remaining bound states we typically see
an enhancement of the mass of the order of $100$--$200$ MeV,
depending upon the meson channel. That is, inclusion of the non-Abelian
diagram is repulsive in the meson channels considered here.

If we turn off the dominant
non-Abelian correction and instead consider the sub-leading Abelian
diagram (AB) we obtain the results labelled AB in
Table~\ref{tab:results}. Based on the relative strength of the
correction due to its colour factors we expect the results to be
$N_c^2$ suppressed, and indeed this is the case.  The mass of the pion
is protected by chiral symmetry and so it receives negligible
contributions from such corrections beyond rainbow-ladder. For the
scalar, vector and $a_1$ axial-vector we see mass reductions of the
order of $20-40$~MeV, while for the $b_1$ there is a slight repulsion of
$3$~MeV. This gives a strong indication that such Abelian corrections to
rainbow-ladder are generally small and attractive.

Now, if we consider both the Abelian and non-Abelian corrections to the
quark-gluon vertex together we might expect the effects on the meson
bound-state to stack. That is, for the vector meson we would expect to see a
$\sim120$ repulsion from the NA correction, with a $\sim 25$ attraction
from the AB correction, resulting in a bound-state mass of $\sim 860$. 
Instead, we see that the Abelian
corrections are heavily suppressed by the non-Abelian ones for all meson channels, giving
results that are almost identical to the those from the NA diagram
alone. The conclusion is that naively adding and subtracting the results 
of different independent studies without performing the combined dynamical 
calculation can be misleading. It is thus reasonable, with the
interaction model presented here, to ignore the Abelian diagram completely 
whenever we take the dominant non-Abelian diagram into account.

Finally, having demonstrated that the Abelian diagram is significantly
suppressed in the meson spectrum, we consider the dominant non-Abelian 
corrections beyond rainbow-ladder with
unquenching effects in the form of pion exchange (PI). This gives rise
to the fifth row of Table~\ref{tab:results}. As expected, the inclusion of a
pion exchange kernel is generally attractive, with an $80$ MeV reduction
of the vector meson mass with respect to the RL+NA result. This
demonstrates, as suspected in~\cite{Bender:1996bb,Fischer:2009jm} the near
cancellation of beyond-the-rainbow corrections with unquenching effects in this
channel. Here, this is not an exact mechanism but the result of
dynamical combinations of attractive and repulsive components of the
quark-gluon vertex; a more accurate picture will be revealed when
improved approximation schemes are employed that permit quantitative
study.

\section{Discussion and Outlook}\label{sec:outlook}
We improved upon previous beyond-the-rainbow investigations, which
focused upon pion unquenching
corrections~\cite{Fischer:2007ze,Fischer:2008sp,Fischer:2008wy} and the
dominant non-Abelian corrections to the quark-gluon
vertex~\cite{Fischer:2009jm} by including the sub-leading Abelian
corrections to the quark-gluon vertex. 

The study we presented involves many technical developments in the
calculation of the quark propagator and quark-gluon vertex at complex
momenta. For highly non-trivial two-loop diagrams such as the
crossed-ladder kernel, new and improved solution methods were required
to render the calculation tractable. Despite these developments, the
study remains qualitative and serves only as an exploratory study of the
relevance of the different contributions beyond rainbow-ladder
considered thus far.

Future investigations will focus upon including realistic
input from the ghost and gluon sector of the theory. These 
are currently under investigation and the results will be reported elsewhere.

\acknowledgments{This work was supported by the Helmholtz-University Young Investigator Grant No. VH-NG-332 and by the Helmholtz International Center for FAIR within the LOEWE program of the State of Hesse.
}

\end{multicols}

\vspace{-2mm}
\centerline{\rule{80mm}{0.1pt}}
\vspace{2mm}

\begin{multicols}{2}

\end{multicols}

\vspace{5mm}

\clearpage

\end{document}